\documentclass[10pt,aps,twocolumn,showpacs,pra]{revtex4}
\usepackage{amssymb,amsfonts,amsmath,wrapfig,mathrsfs,mathbbol}
\usepackage{graphicx}
\begin{document}


\title{Spin-induced angular momentum switching}

\author{Gabriel F. Calvo and Antonio Pic\'{o}n}


\date{\today}

\begin{abstract}
\hspace*{-2.7mm} When light is transmitted through optically inhomogeneous and anisotropic media the spatial distribution of light can be modified according to its input polarization state. A complete analysis of this process, based on the paraxial approximation, is presented, and we show how it can be exploited to produce a spin-controlled-change in the orbital angular momentum of light beams propagating in patterned space-variant-optical-axis phase plates. We also unveil a new effect. The development of a strong modulation in the angular momentum change upon variation of the optical path through the phase plates.
\end{abstract}
\maketitle

In the past few years considerable interest has been attracted to the generation, manipulation and characterization of helical beams. These beams can transport angular momentum ({\em spin} and {\em orbital}) along their propagation direction~\cite{OAM}, producing mechanical effects (optical torques) that have been exploited for trapping atoms, molecules, and macroscopic particles~\cite{He,Paterson,ONeil,Grier}. Other applications include the detection of rotational frequency shifts~\cite{Courtial,Basistiy}, geometric phases~\cite{Padgett99,Galvez,Calvo05}, and for spatial mode encoding~\cite{Vaziri,Gabi04,Padgett04,Calvo06}, demonstrated both at the classical and single photon level. For instance, the possibility of simultaneously using the spin (polarization) and orbital angular momenta of light is becoming increasingly appealing for small-scale quantum information tasks~\cite{Barreiro}. 
\par
The aim of this letter is to address the problem of how to exploit the spin degree of freedom of vectorial helical beams to induce changes in their orbital angular momentum. Prior to coming to the matter, we mention previous important contributions connected with this process. Generation of non-scalar helical waves, based on spatially nonuniform polarization transformations, with subwavelength diffraction gratings in the mid-infrared has been reported~\cite{Hasman,Niv}. A proof-of-principle demonstration of spin-controlled-changes in the orbital angular momentum of circularly-polarized Gaussian beams in the visible domain, using patterned nematic liquid crystals, has been experimentally achieved~\cite{Marrucci}. Also, a method to control the transfer of spin with an externally applied dc electric field in an optically active medium has been proposed~\cite{Chen}. Our approach, based on the vectorial paraxial propagation of helical beams in space-variant-optical-axis media, enables us to describe in a remarkable simple way not only the mechanism of spin-to-orbital angular momentum switching, but also to reveal a new effect: the development of a strong modulation in the spin and orbital angular momentum changes when varying the traversed optical path. Possible uses for polarization-entanglement transfer onto orbital angular momentum-entanglement in two-photon states are also discussed.
\par
\begin{figure}[b]
\begin{center}
\hspace*{-0.2cm}
\hbox{\vbox{\vskip 0mm \includegraphics[width=60mm]{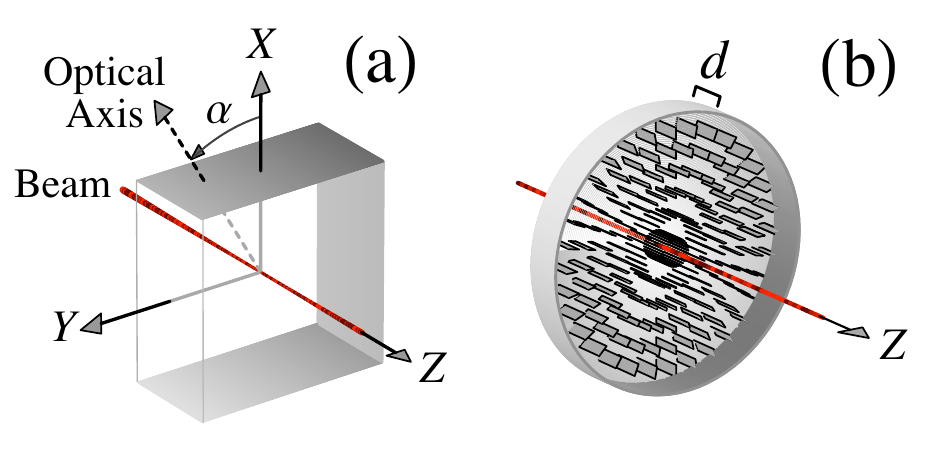}}}
\end{center}
\vspace*{-0.5cm}
\caption{\small (color online) (a) Axes configuration for homogeneous uniaxial media. (b) Azimuthally inhomogeneous uniaxial phase plate. Segments represent the local orientation of the optical axis. Light propagation is along the $Z$ direction.}
\label{fig:Axes}
\end{figure}
Propagation of monochromatic light (of frequency $\omega$) in anisotropic linear media is described, starting from Maxwell's equations, by~\cite{Ciattoni03} $\nabla^{2}{\bf E}-\nabla(\nabla\cdot{\bf E})+k_{0}^{2}\hat{\boldsymbol\epsilon}\cdot{\bf E}=0$. Here, $k_{0}=\omega/c$, $\hat{\boldsymbol\epsilon}$ is the relative dielectric tensor (at frequency $\omega$), and ${\bf E}$ is the complex amplitude of the electric field. We focus our analysis on uniaxial media in which the optical axis is confined in a plane orthogonal to the propagation direction of the incident light (along the $Z$ axis), so that the walk-off effect between the ordinary and extraordinary components of the field is absent. Absorption is neglected. Let us first regard the medium as homogeneous. In the principal axes reference frame the relative dielectric tensor is represented by a diagonal matrix $\hat{\boldsymbol\epsilon}_{p}=\textrm{diag}(n_{e}^{2},n_{o}^{2},n_{o}^{2})$, with $n_{o}$ and $n_{e}$ being the ordinary and extraordinary refractive indexes, respectively. If the optical axis is rotated an angle $\alpha$ about the $Z$ axis with respect to a fixed reference frame [see Fig. \ref{fig:Axes}(a)], the rotated dielectric tensor $\hat{\boldsymbol\epsilon}(\alpha)$ in the $XYZ$ frame is connected with $\hat{\boldsymbol\epsilon}_{p}$ by a similarity transformation $\hat{\boldsymbol\epsilon}(\alpha)=\hat{R}_{z}(\alpha)\hat{\boldsymbol\epsilon}_{p}\hat{R}_{z}(-\alpha)$, where $\hat{R}_{z}(\alpha)$ is the usual $3\times3$ rotation matrix about the $Z$ axis~\cite{rotation}. In order to solve the propagation equation in the fixed frame, we express the field in terms of the standard two-dimensional Fourier integral ${\bf E}({\bf r},z)=\int \textrm{d}^{2}{\bf q}\exp(i{\bf q}\cdot{\bf r})\boldsymbol{\mathcal{E}}({\bf q},z)$, where ${\bf r}=x{\bf u}_{x}+ y{\bf u}_{y}$ and ${\bf q}=k_{x}{\bf u}_{x}+k_{y}{\bf u}_{y}$ are the transverse position and wavevector components. We look for plane-wave solutions $\boldsymbol{\mathcal{E}}({\bf q},z)=\boldsymbol{\mathcal{E}}_{o}({\bf q})\exp(ik_{oz}z)+\boldsymbol{\mathcal{E}}_{e}({\bf q})\exp(ik_{ez}z)$, with $k_{oz}=[k_{0}^{2}n_{o}^{2}-q^{2}]^{1/2}$ and $k_{ez}=[k_{0}^{2}n_{e}^{2}-(k_{x}\cos\alpha+k_{y}\sin\alpha)^{2}n_{e}^{2}/n_{o}^{2}-(k_{x}\sin\alpha-k_{y}\cos\alpha)^{2}]^{1/2}$. The ordinary and extraordinary amplitudes $\boldsymbol{\mathcal{E}}_{o,e}$ can be explicitly obtained from the boundary condition of the field, $\boldsymbol{\mathcal{E}}_{0}({\bf q})$, at $z=0$. By resorting to the paraxial approximation, which amounts to retaining only the low spatial transverse frequencies ($\vert {\bf q}\vert\ll k_{0}$), the relevant contribution of the angular spectrum of the field is found to be given by its transverse part $\boldsymbol{\mathcal{E}}_{\perp}({\bf q},z)=\hat{U}_{\alpha}({\bf q},z)\boldsymbol{\mathcal{E}}_{\perp0}({\bf q})$, where
\begin{eqnarray}
\hat{U}_{\alpha}({\bf q},z)\!\!\!\!&=&\!\!\!\! \frac{\exp(ik_{ez}z)+\exp(ik_{oz}z)}{2}\hat{\mathbb{1}}\nonumber\\
\!\!\!\!&+&\!\!\!\!\frac{\exp(ik_{ez}z)-\exp(ik_{oz}z)}{2}\hat{\mathcal{R}}_{z}(\alpha)\hat{\sigma}_{z}\hat{\mathcal{R}}_{z}(-\alpha)\,, \nonumber
\end{eqnarray}
$\hat{\mathcal{R}}_{z}(\alpha)$ and $\hat{\sigma}_{z}$ denoting the $2\times2$ rotation and the Pauli matrices about the $Z$ axis, respectively.
\par
Let us now examine the more general situation in which the orientation $\alpha$ of the optical axis in the $XY$ plane varies with position [remaining uniform along the $Z$ direction, see Fig.~\ref{fig:Axes}(b)]. This is of relevance for patterned nematic liquid crystal phase plates~\cite{Marrucci}. Let $d$ denote their thickness. To obtain the output field in this situation we transform back $\boldsymbol{\mathcal{E}}_{\perp}$ to real space variables and integrate $\hat{U}_{\alpha}$ over the transverse spatial frequencies. This is an excellent approximation as long as $\alpha$ varies smoothly on the wavelength scale of the input beam. The transverse field at the output face of the phase plates is 
\begin{eqnarray}
{\bf E}_{\perp}(r,\phi,d)= \frac{k_{0}n_{o}}{2\pi id}\int_{0}^{\infty}\!\!\int_{0}^{2\pi}\rho\,\textrm{d}\rho\, \textrm{d}\varphi\left\{\left[\frac{F_{e}+F_{o}}{2}\right]\hat{\mathbb{1}}\right.\nonumber\\
+ \left. \left[\frac{F_{e}-F_{o}}{2}\right]\hat{\mathcal{R}}_{z}(\alpha)\hat{\sigma}_{z}\hat{\mathcal{R}}_{z}(-\alpha)\right\}
{\bf E}_{\perp0}(\rho,\varphi)\, , \label{eq:RSField}
\end{eqnarray}
where $F_{o}=\exp\left[ik_{0}n_{o}d+i\beta_{o}\vert {\bf r}-{\boldsymbol\rho}\vert^{2}\right]$ is the ordinary Fresnel kernel with $\vert {\bf r}-{\boldsymbol\rho}\vert^{2}=r^{2}+\rho^{2}-2r\rho\cos(\phi-\varphi)$, $\beta_{o}=k_{0}n_{o}/2d$, and the input transverse field of the incident beam is ${\bf E}_{\perp0}$. The extraordinary Fresnel kernel reads
\begin{eqnarray}
F_{e}=\exp\left\{\!ik_{0}n_{e}d+i\beta_{e}\vert {\bf r}-{\boldsymbol\rho}\vert^{2}+i\Delta\beta_{e}\!\left[r^{2}\!\cos2(\alpha-\phi)\right.\right.\nonumber\\
+\left.\left. \rho^{2}\!\cos2(\alpha-\varphi)-2r\rho\cos(2\alpha-\phi-\varphi)\right]\right\}\! , \nonumber
\end{eqnarray}
with $\beta_{e}=k_{0}(n_{o}^{2}+n_{e}^{2})/4n_{e}d$ and $\Delta\beta_{e}=k_{0}(n_{o}^{2}-n_{e}^{2})/4n_{e}d$. Equation~(\ref{eq:RSField}) provides our first main result. It shows that the action of any uniaxial phase element with space-constant- or space-variant-optical-axis oriented orthogonally to the propagation of light produces two distinct field contributions: i) a term that preserves the polarization state of the input field (the one proportional to the identity matrix $\hat{\mathbb{1}}$), and ii) a term that rotates the polarization state of the input field. The spatial distribution of the input field is modified in both cases, although in a different fashion. The first term is always present and dominates over the second term. Notice that Eq.~(\ref{eq:RSField}) reduces to the well-known isotropic Fresnel integral when $n_{o}=n_{e}$ (the second term vanishes). The two contributions can however be separated by using a suitable combination of polarizers and a Mach-Zender interferometer. For input circular polarization the two output contributions have (opposite) circular polarization and no interferometers are required; one could employ the recently proposed chiral liquid microcells that enable the splitting of circularly-polarized beams~\cite{Ghosh}. When the birefringence is not too large ($\vert\Delta\beta_{e}\vert\ll\beta_{e}$), it is possible to obtain an accurate approximation for the extraordinary Fresnel kernel $F_{e}\simeq\exp\left[ik_{0}n_{e}d+i\beta_{e}\vert {\bf r}-{\boldsymbol\rho}\vert^{2}\right]$. The neglected part is responsible for a small astigmatism in the extraordinary part of the field profile. We use this approximation henceforth.
\par
Consider now a normalized input transverse field of the form ${\bf E}_{\perp0}(\rho,\varphi)= [a{\bf u}_{x}+ b{\bf u}_{y}]LG_{\ell,p}(\rho)\exp(i\ell\varphi)$, with degree of polarization $\sigma=i(ab^{*}-a^{*}b)$. Here $\sigma=\pm1$ for left- and right-hand circularly polarized light, respectively, whereas $\sigma=0$ for linearly polarized. The functions $LG_{\ell,p}(\rho)$ denote the Laguerre-Gaussian modes~\cite{OAM,Calvo06}; indices standing for topological charge $\ell=0,\pm1,\pm2,\ldots$, and non-axial radial node number $p=0,1,2,\ldots$. The properties of these helical beams expressed in a cylindrically polarization basis under strong focusing conditions have been recently studied~\cite{Zhan}. Within the paraxial approximation, the total optical angular momentum (integrated over the transverse plane) along the propagation direction of the input beam can be decomposed as~\cite{OAM} $J_{z}=L_{z}+S_{z}$, where $L_{z}=\ell/\omega$ and $S_{z}=\sigma/\omega$ correspond to the orbital and spin angular momenta per unit energy.
\par
We wish to determine whether, according to the spin state of the above input beam, the total angular momentum $J_{z}$ is conserved or changed upon propagation through space-variant-optical-axis phase plates. In the general case where the transverse electric field, ${\bf E}_{\perp}=v_{x}{\bf u}_{x}+ v_{y}{\bf u}_{y}$, consists of two distinct spatial distributions (position-dependent polarization), $v_{x}$ and $v_{y}$, the output orbital and spin angular momenta can be cast as
\begin{eqnarray}
L_{z}\!\!\!\!&=&\!\!\!\!\frac{i}{2\omega}\!\sum_{j=x,y}\int_{0}^{\infty}\!\!\!\int_{0}^{2\pi}\!r\,\textrm{d}r\,\textrm{d}\phi\left[ v_{j}\frac{\partial v^{*}_{j}}{\partial\phi}- v^{*}_{j}\frac{\partial v_{j}}{\partial\phi}\right]\!,\label{eq:Lz}\\
S_{z}\!\!\!\!&=&\!\!\!\!\frac{i}{2\omega}\!\int_{0}^{\infty}\!\!\!\int_{0}^{2\pi}\!r^{2}\,\textrm{d}r\,\textrm{d}\phi\, \frac{\partial}{\partial r}\left[v^{*}_{x}v_{y}-v_{x}v^{*}_{y}\right]\!.\label{eq:Sz} 
\end{eqnarray}
\par
Assume that the orientation of the optical axis in the phase plates is described by the following azimuthal relation~\cite{Marrucci} $\alpha(\varphi)=q\varphi + \alpha_{0}$, with constants $q$ and $\alpha_{0}$. Only if $q$ is an integer or a semi-integer the optical axis does not possess discontinuity lines in the phase plates, but only a defect in their center. For input beams corresponding to a fundamental Gaussian mode ($\ell=p=0$), the field at the output face of the plates will generally exhibit an abrupt steepening in the vicinity of the origin. 
\par
Remarkably, when $v_{x}$ and $v_{y}$ represent the field components in Eq.~(\ref{eq:RSField}), one can integrate Eqs.~(\ref{eq:Lz}) and (\ref{eq:Sz}) for all Laguerre-Gaussian modes. They yield an strikingly simple formula for the total angular momentum change (per unit energy) $\Delta J_{z}=\Delta L_{z}+\Delta S_{z}$, where the changes in the orbital and spin parts are given by
\begin{eqnarray}
\Delta L_{z}\!\!\!\!&=&\!\!\!\!\frac{\sigma q}{4\pi\omega}\left[1+\frac{\beta_{o}^{2}}{\beta_{e}^{2}}-\frac{2\beta_{o}}{\beta_{e}}\cos\left(k_{0}\vert n_{o}-n_{e}\vert d\right)\right]\!,
\label{eq:Lzlp}\\
\Delta S_{z}\!\!\!\!&=&\!\!\!\!-\frac{\sigma}{4\pi\omega}\left[1+\frac{\beta_{o}^{2}}{\beta_{e}^{2}}-\frac{2\beta_{o}}{\beta_{e}}\cos\left(k_{0}\vert n_{o}-n_{e}\vert d\right)\right]\!.
\label{eq:Szlp}
\end{eqnarray}
Notice that none of the angular momentum parts depend on the specific input Laguerre-Gaussian modes nor on the beam width. If $q=1$, then $\Delta J_{z}=0$, irrespective of the input spin. This fact does not preclude that the orbital and spin $L_{z}$ and $S_{z}$ exchange their magnitudes (equal and of opposite sign). The experiment reported in Ref.~cite{Marrucci} referred to the case of fundamental Gaussian modes as input beams and plates with $q=1$. There, it was confirmed that the initial total angular momentum was conserved, in complete agreement with Eqs.~(\ref{eq:Lzlp}) and (\ref{eq:Szlp}). This fact is expected in view of the full cylindrical symmetry of both the light profile and the plates [see Fig.~\ref{fig:Axes}(b)], so that the total angular momentum carried by the incident beam must be a conserved quantity via Noether's theorem. Interestingly enough, one sees another effect from Eqs.~(\ref{eq:Lzlp}) and (\ref{eq:Szlp}). Namely, that the orbital, spin, and total angular momentum change show the development of a very strong modulation when varying the thickness $d$ of the phase plates. By slightly changing the incidence angle of the input beam on the plates (or by tilting the plates, in a similar fashion as the well-known Maker fringes are detected when measuring the efficiency in second harmonic generation), this effect should be observable as a net transfer of angular momentum to absorbing particles~\cite{OAM} trapped on the beam axis in an optical tweezer~\cite{He,Paterson}. An off-axis geometry~\cite{ONeil} would also reveal both spin and orbital components. For fixed $q\neq1$, the maximum change of angular momentum occurs when the input beams are circularly polarized ($\sigma=\pm1$) and the anisotropic plate thicknesses are $d=(2m+1)\lambda/2\vert n_{o}-n_{e}\vert$, $m=0,1,2,\ldots$, that is, of the order of the beam wavelength $\lambda$. Moreover, if the input beams are linearly polarized ($\sigma=0$), $\Delta J_{z}=0$ holds, and, in particular, the orbital part (for Laguerre-Gaussian modes with topological charge $\ell$) is thus conserved ($L_{z}=\ell/\omega$). 
\par
In summary, from the above results it is manifest that a finite exchange of angular momentum between the input beam and the space-variant-optical-axis phase plates will generally take place. However, this transfer will depend both on the optical spin and the $q$-parameter of the plate, but not on the particular spatial light mode. In this respect, it would also be quite interesting to identify other physical scenarios, {\em complementary} to the one described here, in which the change of the distinct optical angular momentum parts could solely be mediated by the input orbital angular momentum but not on the spin. The combined action of such two complementary systems would enable one to perform full-fledge controlled-switching between the two angular momenta degrees of freedom, of great relevance for quantum information processing based on linear optical schemes~\cite{Walther}. For instance, it should then be possible to transfer two-photon entanglement~\cite{Calvo07}, either in spin or in orbital angular momentum, onto the other degree of freedom, a necessary operation for the so-called quantum repeaters, where the interface between a quantum communication channel and a quantum memory will probably require manipulation of entanglement involving several degrees of freedom of light.
\par
We thank fruitful discussions with M.\,J. Padgett and acknowledge financial support from the Spanish Ministry of Science and Technology through Project No. FIS2005-01369, the Juan de la Cierva Grant Program, and the CONSOLIDER2006-00019 Program.
\par


\begin{thebibliography}{99}

\bibitem{OAM} L. Allen, S.\,M. Barnett, and M.\,J. Padgett, {\em Optical Angular Momentum} (IOP Publishing, Bristol, UK, 2003).

\bibitem{He} H. He, M.\,E.\,J. Friese, N.\,R. Heckenberg, and H. Rubinsztein-Dunlop, Phys. Rev. Lett. {\bf 75}, 826 (1995).

\bibitem{Paterson} L. Paterson, M.\,P. MacDonald, J. Arlt, W. Sibbett, P.\,E. Bryant, and K. Dholakia, Science {\bf 292}, 912 (2001).

\bibitem{ONeil} A.\,T. O'Neil, I. MacVicar, L. Allen, and M.\,J. Padgett, Phys. Rev. Lett. {\bf 88}, 053601 (2002).

\bibitem{Grier} D.\,G. Grier, Nature (London) {\bf 424}, 810 (2003).

\bibitem{Courtial} J. Courtial, K. Dholakia, L. Allen, and M.\,J. Padgett, Phys. Rev. Lett. {\bf 81}, 4828 (1998).

\bibitem{Basistiy} I.\,V. Basistiy, V.\,V. Slyusar, M.\,S. Soskin, M.\,V. Vasnetsov, and A.\,Ya. Bekshaev, Opt. Lett. {\bf 28}, 1185 (2003).

\bibitem{Padgett99} M. J. Padgett and J. Courtial, Opt. Lett. {\bf 24}, 430 (1999).

\bibitem{Galvez} E.\,J. Galvez, P.\,R. Crawford, H.\,I. Sztul, M.\,J. Pysher, P.\,J. Haglin, and R.\,E. Williams, Phys. Rev. Lett. {\bf 89}, 203901 (2003).

\bibitem{Calvo05} G.\,F. Calvo, Opt. Lett. {\bf 30}, 1207 (2005).

\bibitem{Vaziri} A. Vaziri, J.-W. Pan, T. Jennewein, G. Weihs, and A. Zeilinger, Phys. Rev. Lett. {\bf 91}, 227902 (2003).

\bibitem{Gabi04} G. Molina-Terriza, A. Vaziri, J. \v{R}eh\'{a}\v{c}ek, Z. Hradil, and A. Zeilinger, Phys. Rev. Lett. {\bf 92}, 167903 (2004).

\bibitem{Padgett04} G. Gibson, J. Courtial, M.\,J. Padgett, M. Vasnetsov, V. Pas'ko, S.\,M. Barnett, and S. Franke-Arnold, Opt. Express {\bf 12}, 5448 (2004).

\bibitem{Calvo06} G.\,F. Calvo, A. Pic\'{o}n, and E. Bagan, Phys. Rev. A {\bf 73}, 013805 (2006).

\bibitem{Barreiro} J.T. Barreiro, N.K. Langford, N.A. Peters, and P.G. Kwiat, Phys. Rev. Lett. {\bf 95}, 260501 (2005).

\bibitem{Hasman} E. Hasman, G. Biener, A. Niv, and V. Kleiner, in {\em Progress in Optics}, E. Wolf, ed. (North-Holland, 2005), Vol. 47, pp. 215-289.

\bibitem{Niv} A. Niv, G. Biener, V. Kleiner, and E. Hasman, Opt. Lett. {\bf 30}, 2933 (2005).

\bibitem{Marrucci} L. Marrucci, C. Manzo, and D. Paparo, Phys. Rev. Lett. {\bf 96}, 163905 (2006).

\bibitem{Chen} L. Chen, G. Zheng, J. Xu, B. Zhang, and W. She, Opt. Lett. {\bf 31}, 3474 (2006).

\bibitem{Ciattoni03} A. Ciattoni and C. Palma, J. Opt. Soc. Am. A {\bf 20}, 2163 (2003).

\bibitem{rotation} The rotation matrix is given by 
\begin{eqnarray}
\hat{R}_{z}(\alpha)=\left[\!\begin{array}{ccc}
\cos\alpha&-\sin\alpha&0
\\
 \sin\alpha&\cos\alpha&0
\\
0&0&1\end{array}\!\right]\, . \nonumber
\end{eqnarray}

\bibitem{Ghosh} A. Ghosh and P. Fischer, Phys. Rev. Lett. {\bf 97}, 173002 (2006).

\bibitem{Zhan} Q. Zhan, Opt. Lett. {\bf 31}, 867 (2006).

\bibitem{Walther} P. Walther, K.\,J. Resch, T. Rudolph, E. Schenck, H. Weinfurter, V. Vedral, M. Aspelmeyer, and A. Zeilinger, Nature (London) {\bf 434}, 169 (2005).

\bibitem{Calvo07} G.\,F. Calvo, A. Pic\'{o}n, and A. Bramon, Phys. Rev. A {\bf 75}, 012319 (2007).

\end{thebibliography}
\end{document}